\documentclass[%
reprint,
superscriptaddress,
showpacs,preprintnumbers,
showkeys,
 amsmath,amssymb,
 aps,
]{revtex4-1}

\usepackage{graphicx}% Include figure files
\usepackage{dcolumn}% Align table columns on decimal point
\usepackage{bm}% bold math
\def\be{\begin{equation}}
\def\ee{\end{equation}}
\def\bea{\begin{eqnarray}}
\def\eea{\end{eqnarray}}
\newcommand{\ket}[1]{\mbox{$|#1\rangle$}}
\newcommand{\bra}[1]{\mbox{$\langle#1|$}}

\begin{document}

\title {Quantum Information Transfer between Topological and Superconducting  Qubits}

\author{Fang-Yu Hong}
\email[Email address:]{honghfy@163.com}
\email[Tel:]{86-571-86843468}
\affiliation{Department of Physics, Center for Optoelectronics Materials and Devices, Zhejiang Sci-Tech University,  Hangzhou, Zhejiang 310018, China}
%\author{Shi-Jie Xiong}
%\affiliation{National Laboratory of Solid State Microstructures and
%Department of Physics, Nanjing University, Nanjing 210093, China}
\author{Jing-Li Fu}
\affiliation{Department of Physics, Center for Optoelectronics Materials and Devices, Zhejiang Sci-Tech University,  Hangzhou, Zhejiang 310018, China}
%\author{Yang Xiang}
%\affiliation{School of Physics and Electronics, Henan University, Kaifeng, Henan 475004, China}
%\author{W.H. Tang}
%\affiliation{Department of Physics, Center for Optoelectronics Materials and Devices, Zhejiang Sci-Tech University,  Hangzhou, Zhejiang 310018, China}
\author{Zhi-Yan Zhu}
\affiliation{Department of Physics, Center for Optoelectronics Materials and Devices, Zhejiang Sci-Tech University,  Hangzhou, Zhejiang 310018, China}
%\author{Li-zhen Jiang}
%\affiliation{College of Information and  Electronic Engineering, Zhejiang Gongshang University, Hangzhou, Zhejiang 310018,China}
%\author{Liang-neng Wu}
%\affiliation{College of Science, China Jiliang University, Hangzhou, Zhejiang 310018, China}
\date{\today}
\begin{abstract}
 We describe a scheme that enables a strong Jaynes-Cummings coupling between a topological qubit and a superconducting flux qubit. The coupling strength is dependent on  the  phase difference between two superconductors on a topological insulator and may be expediently controlled by a phase controller. With this coherent coupling and single-qubit rotations arbitrary unitary operations  on  the two-qubit  hybrid system of  topological and flux qubits can be performed. Numerical simulations show that quantum state transfer and  entanglement distributing between the topological and superconducting flux qubits may be performed with high fidelity.
\end{abstract}


\maketitle
\section{INTRODUCTION}
 The decoherence of quantum states by the environment is the main obstacle in the way towards realizing quantum computers. To circumvent this difficulty some interesting topological quantum computation schemes \cite{ayki,cnsh} have been suggested, where quantum information is stored in nonlocal (topological) degrees of freedom of topologically ordered systems. These nonlocal degrees of freedom are decoupled from local perturbations, enabling the topological approach to quantum information processing to obtain its exceptional fault tolerance and to have a tremendous advantage over conventional ones.  The simplest non-Abelian excitation for topological qubits is the zero energy Majorana bound state (MBS) \cite{fwil}, which is predicted to be  exist in the spin lattice systems \cite{ayki}, in the $p+ip$ superconductors \cite{ nrdg}, in the filling fraction $\nu=5/2$ fractional quantum Hall system \cite{cnsh}, in the superconductor Sr$_2$RuO$_4$ \cite{sscn}, in the   topological insulators \cite{lfck,mhck}, and in some semiconductors with strong spin-orbit interaction \cite{jsrl,jali,yogr,rljs, vmou}.

However, the local decoupling makes measuring and manipulating  topological states difficult because they can only be manipulated by globe braiding operations, i.e., by physical exchange of the associated local quasiparticle non-Abelian excitations \cite{aste,daiv}. Moreover topologically protected braiding operations for Ising anyons alone are  not adequate to fulfill  universal quantum computation and have to be supplemented with topologically unprotected operations \cite{pbon,pbond}. Within a topological system unprotected operations prove to be very challenging because of  significant nonuniversal effects \cite{pbrl}. On the other hand, conventional quantum information processing systems have been advancing steadily, such as the recent progresses in quantum network using single atoms in optical cavities \cite{srcn}, in long coherence times of nuclear spins in a diamond crystal \cite{pmgk,mglc}, in  high fidelity manipulations on trapped ions \cite{rbdw} and on superconducting qubits \cite{jcfw}, in generation of  entanglement between single-atoms at a distance \cite{dmpm} and between a photon and a solid-state spin qubit \cite{etyc}

  Thus it is highly desirable to combine the advantages of conventional qubits  with those of topological qubits to construct hybrid systems, where the necessary topologically unprotected gates can be imported from the conventional quantum systems (CQS) and  topological states can be transferred to CQS for high fidelity readout. Such hybrid systems have been considered recently  for the anyons in optical lattices \cite{ljia,magu} and for the Majorana anyons coupled to superconducting flux qubits \cite{fhaa,jsst,ljck} or to a semiconductor double-dot qubit \cite{pbrl}.

Here we propose a scheme for quantum information transfer between a superconducting  flux qubit \cite{jemo,zxss,jcfw} and a topological qubit encoded on  Majorana fermions (MFs) at the junctions among three superconductors mediated by a topological insulator (TI) \cite{lfck}. The strong Jaynes-Cummings (JC) coupling between  topological and  superconducting flux qubits can be obtained on the basis of the interaction between two MFs located at the two ends of a  linear superconductor-TI-superconductor (STIS) junction, and be coherently controlled by the phase differences between the two superconductors of the STIS junction. With this strong coupling at hand, arbitrary quantum information transfer and quantum entanglement distribution between the topological and  the flux qubits can be accomplished with near unit fidelity.

\begin{figure}[t]
\includegraphics[width=8cm]{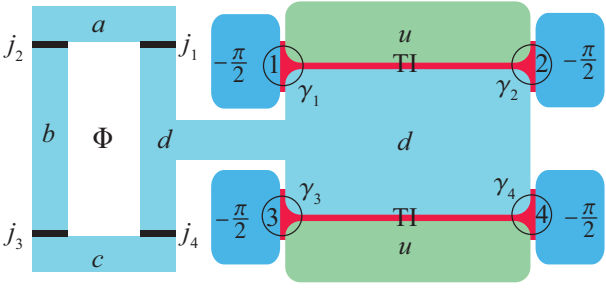}
\caption{\label{fig1}(color online). Schematics for a hybrid system of topological and superconducting flux qubits. A flux qubit is made up of four Josephson junctions ($j_{1,2,3,4}$) and four superconducting islands ($a,b,c,d$) patterned on the surface of a topological insulator, enclosing an external flux $\Phi\approx h/4e$. A topological qubit consists of two pairs of Majorana fermions ($(\gamma_1,\gamma_2)$ and $(\gamma_3,\gamma_4)$). Island $d$ is shared by the topological and the flux qubits. Two Majorana fermions (marked with circles) at two superconducting trijunctions are coupled though STIS quantum wire with coupling strength dependent on the phase $\phi_d$ of island $d$ relative to $\phi_u=-\pi$.    }
\end{figure}

 \section {Hybrid system}
 The prototype hybrid  quantum system  shown in Fig.1  is made up of a superconducting flux qubit  and a topological qubit encoded on four MFs.  The flux qubit consists of a loop of  four Josephson junctions  ($j_{1,2,3,4}$) and four superconducting island $a,b,c,d$, enclosing an externally applied magnetic flux $\Phi\approx\frac{h}{4e}$. The MFs are described by Majorana fermion operators $\gamma_i(i=1,2,3,4)$, which are self-Hermitian, $\gamma_i^\dagger=\gamma_i$, and fulfill fermionic anticommutation relation $\{\gamma_i,\gamma_j\}=\delta_{ij}$.  The Majorana fermion  $\gamma_i$ is localized  at  trijunction $i(i=1,2,3,4)$, which comprises three superconductors divided by a TI \cite{lfck}.
A pair of MFs operators $\gamma_i,\gamma_j$ connected by a STIS wire of length $L$  can form a Dirac fermion operator $f_{ij}=(\gamma_i-i\gamma_j)/\sqrt{2}$, which creates a fermion and $f_{ij}^\dagger f_{ij}=n_{ij}=0,1$ describes the occupation of the corresponding state. Combining two such fermion states gives the two logical states of the topological qubit  $\ket{0}_t=\ket{0_{12}0_{34}}$ and $\ket{1}_t=\ket{1_{12}1_{34}}$.

The flux qubit is made up of  four Josephson junctions  with Josephson coupling energy $E_{J,1}=E_{J,2}=E_J$, $E_{J,3}=\alpha E_J$, and $E_{J,4}=\beta E_J$, where $0.5<\alpha<1$ and $\beta\gg1$.  For these parameters and an externally applied flux $\Phi=h/4e$, the system has two stable states $\ket{0}_f$ and $\ket{1}_f$ for the flux qubit. Corresponding to these two states there are persistent circulating currents of opposite direction with the corresponding superconducting phase
$\phi_d=\phi_c +\sigma_f^z\theta+\zeta\frac{a+a^\dagger}{\sqrt{2}}$ of island $d$  \cite{ljck}, where $\sigma_f^z=(\ket{0}\bra{0}-\ket{1}\bra{1})_f$, $\theta=\frac{\sqrt{4\alpha^2-1}}{2\alpha\beta}$ is the phase difference across Josephson junction $j_4$, $a$ is the annihilation operator for the flux qubit, $\zeta=(\frac{8E_C}{E_J})^{\frac{1}{4}}\beta^{-\frac{1}{2}}$ is the magnitude of quantum fluctuations, and the phase $\phi_c$ of island $c$  is fixed relative to the phase $\phi_u=-\pi$ of island $u$ by a  phase controller \cite{ljck,ljclk}.

The Hamiltonian for the hybrid system can be written in the form ($\hbar=1$) $H=a^\dagger a\omega_f-\frac{1}{2}E(\phi_d)\sigma_t^z $,
   where $\omega_f=\sqrt{8E_JE_C}$, $\sigma^z_{t}=(\ket{0}\bra{0}-\ket{1}\bra{1})_t$, and the coupling strength $E(\phi_d)$ has the approximate form \cite{ljck}
\be\label{eq2}
E(\phi_d)\approx-1.9(\Lambda_{\phi_d}-0.5)v_F/L\quad \text{ for} \,\, \Lambda_{\phi_d}\leq -5
 \ee
  and
 \be\label{eq3}
 E(\phi_d)\approx2\Delta_0\sin\frac{\phi_d}{2} e^{-\Lambda_{\phi_d}}\sim0 \quad \text{ for}\quad \Lambda_{\phi_d}\gg1,
 \ee
  where $ \Lambda_{\phi_d}\equiv\frac{\Delta_0L}{v_F}\sin\frac{\phi_d}{2}$ with  the effective Fermi velocity $v_F$ and the proximity induced superconducting gap $\Delta_0$.

\begin{figure}[t]
\includegraphics[width=8cm]{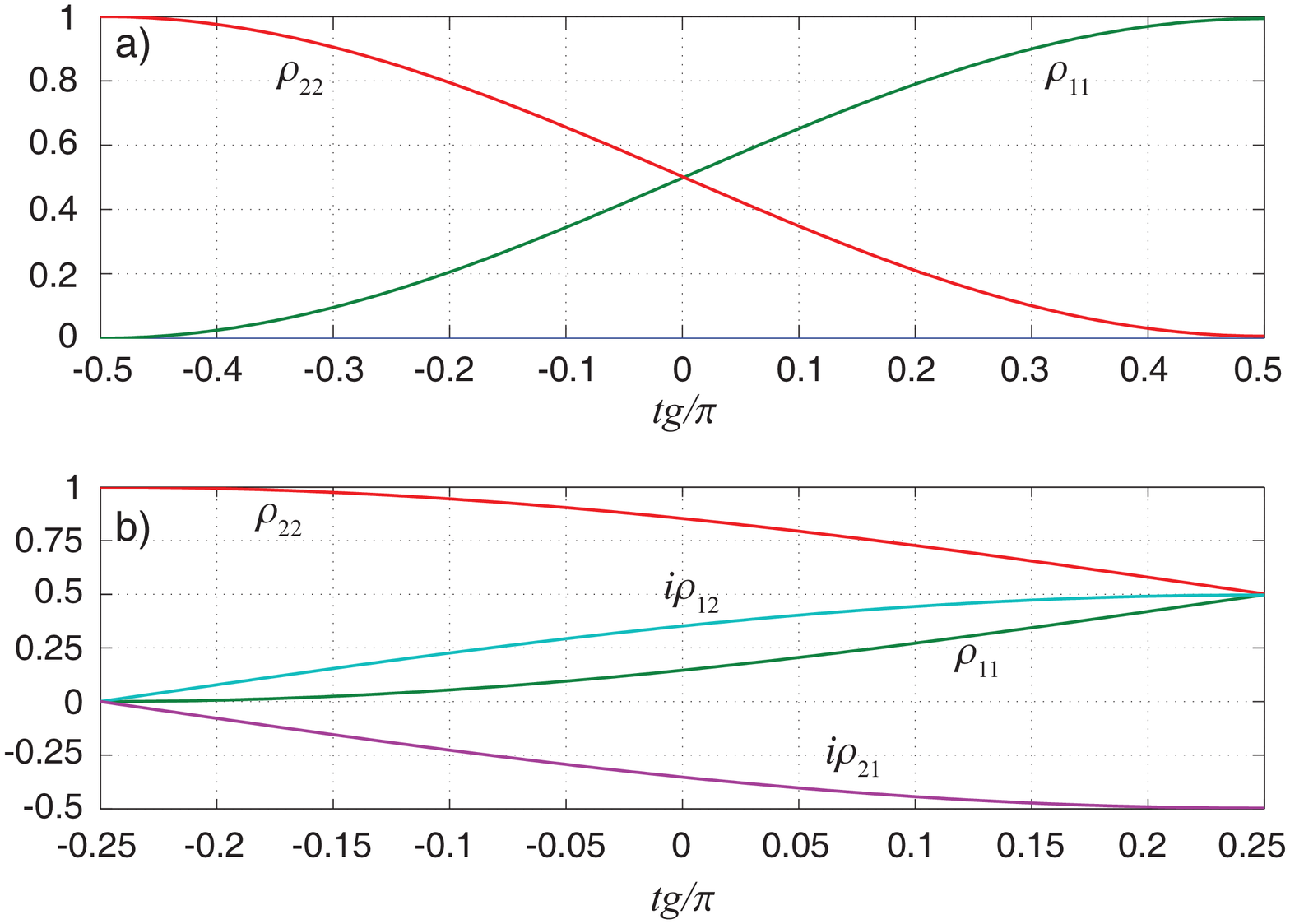}
\caption{\label{fig2}(color online). a) Numerical simulation of the process of the state transfer,  $\ket{\uparrow0}\rightarrow-i\ket{\downarrow1}$. The state transfer fidelity is $F_1=0.993$.  b) Numerical simulation of quantum entanglement generating, $\ket{\uparrow0}\rightarrow(\ket{\uparrow0}-i\ket{\downarrow1})/\sqrt{2}$. The generated  entanglement has a fidelity $F_2=0.996$.  The parameters used are $g/2\pi=-2$GHz, $g'/2\pi=-1$GHz, $T_{f,1}=900$ns, $T_{f,2}=20$ns, and $\omega_f/2\pi=E(\phi_{on})/2\pi=50$ GHz.  The corresponding matrix elements of the density matrix $\rho$ of the hybrid system  are $\rho_{11}=\bra{\downarrow1}\rho\ket{\downarrow1}$, $\rho_{22}=\bra{\uparrow0}\rho\ket{\uparrow0}$, $\rho_{21}=\bra{\uparrow0}\rho\ket{\downarrow1}$, $\rho_{12}=\bra{\downarrow1}\rho\ket{\uparrow0}$.}
\end{figure}

Expanding the coupling strength $E(\phi_d)$ to first order in the small parameters $\frac{\theta}{\omega_f}\frac{dE(\phi)}{d\phi}|_{\phi=\phi_c}$ and $\frac{\zeta}{\omega_f}\frac{dE(\phi)}{d\phi}|_{\phi=\phi_c}$ gives the Hamiltonian
\be \label{eq1}
H=a^\dagger a\omega_f-\frac{1}{2}E(\phi_c)\sigma_t^z-\frac{g'}{2}\sigma_f^z \sigma_t^z-\frac{1}{2}g(a^\dagger+a)\sigma_t^z,
\ee
where
\bea\label{eq10}
g&=&\left.\frac{\zeta}{\sqrt{2}}\frac{dE(\phi)}{d\phi}\right|_{\phi=\phi_c}\notag \\  g'&=&\left.\theta\frac{dE(\phi)}{d\phi}\right|_{\phi=\phi_c}.
\eea
By rewriting Hamiltonian (\ref{eq1}) in terms of $\ket{\downarrow}=\frac{1}{\sqrt{2}}(\ket{0}+\ket{1})_t$ and  $\ket{\uparrow}=\frac{1}{\sqrt{2}}(\ket{0}-\ket{1})_t$ and applying the rotating-wave approximation and the interaction picture we obtain
\bea \label{eq6}
H_I&=&-\frac{1}{2}g(a^\dagger\sigma_t^-+a\sigma_t^+)-\frac{g'}{2}\sigma_f^z (\sigma_t^+e^{iE(\phi_c)t}\notag\\
&+&\sigma_t^-e^{-iE(\phi_c)t}),
\eea
 where $\sigma_t^+=\ket{\uparrow}\bra{\downarrow}$ and $\sigma_t^-=\ket{\downarrow}\bra{\uparrow}$ are the raising and lowering operators, respectively, and  the resonance condition $\omega_f=E(\phi_c)$ has been assumed for simplicity.

 {\it Discussion.}---The first term in $H_I$ \eqref{eq6} describes the JC coupling between the topological and the flux qubits, which is just what we want. The last term will cause the total number of the excitations in the hybrid system changes and will contaminate the quantum information transfer fidelity, thus we may contain its influence by the conditions
 \be\label{eq11}
 g/g'=\frac{\sqrt{2\beta}\alpha}{\sqrt{4\alpha^2-1}}(\frac{8E_C}{E_J})^{\frac{1}{4}}\gg1
  \ee
  and  $E(\phi_c)/g\gg1$. However, because of the factors $e^{\pm iE(\phi_c)t}$ the influence of this non-JC term is very limited, even for the case $g<g'$, which is shown in the following numerical simulation.

According to Eqs.(\ref{eq2}, \ref{eq3}) the JC coupling strength $g$ can be coherently controlled: $g\sim0$ if $\phi_c$ is tuned to $\phi_{\text{off}}$ satisfying $\frac{\Delta_0L}{v_F}\sin\frac{\phi_{\text{off}}}{2} \gg1$, and $g\approx-\Delta_0\frac{\zeta}{\sqrt{2}}\cos\frac{\phi_{\text{on}}}{2}$  if $\phi_c$ is adiabatically adjusted to $\phi_{\text{on}}$ satisfying $\frac{\Delta_0L}{v_F}\sin\frac{\phi_{\text{on}}}{2}\leq-5$. By adiabatically turn on the coupling for a duration corresponding to a $\pi$ pulse $\int g(t)dt=-\pi$, we can perform a unitary transformation
\be
\mu\ket{\downarrow0}+\nu\ket{\uparrow0}\rightarrow \mu\ket{\downarrow0}-i\nu\ket{\downarrow1},
\ee
 accomplishing a quantum state transfer from the topological qubit to the flux qubit by following a single-qubit rotation on the latter, where $\mu$ and $\nu$ are arbitrary complex numbers satisfying $|\mu|^2+|\nu|^2=1$. If we choose $\int g(t)dt=-\pi/2$, we can generate a maximally entangled state $\ket{\uparrow0}\rightarrow(\ket{\uparrow0}-i\ket{\downarrow1})/\sqrt{2}$. Up to a single-qubit rotation a $\sqrt{\text{SWAP}}$ gate, the squared root of SWAP gate, can be obtained by
choosing $\int g(t)dt=-3\pi/2$. With    $\sqrt{\text{SWAP}}$ gates and single-qubit $90^\circ$ rotation about $\hat{z}$ denoted by $\text{R}_z(90)$,  we can obtain the controlled-phase ($\text{CP}_{t,f}$) gate
\be
\text{CP}_{t,f}=\text{R}_{z,t}(90)\text{R}_{z,f}(-90)\sqrt{\text{SWAP}} \text{R}_{z,t}(180) \sqrt{\text{SWAP}}
 \ee
 for the hybrid system.    With  $\text{CP}_{t,f}$ gates and single-qubit rotations an arbitrary unitary transformation on the hybrid system is available \cite{mnic}.
\begin{figure}[t]
\includegraphics[width=8cm]{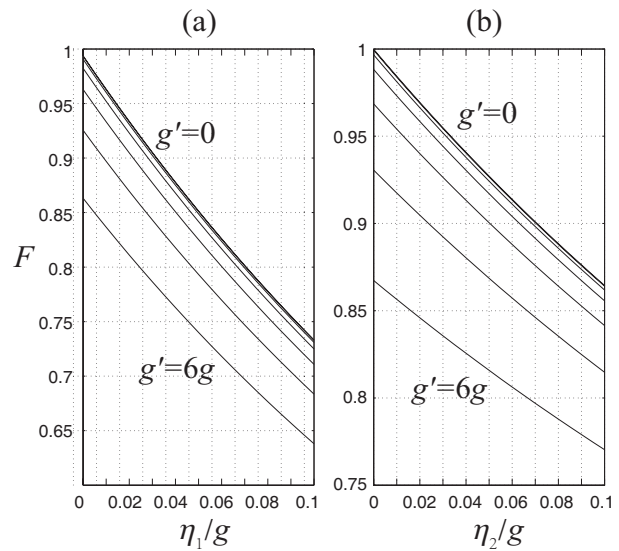}
\caption{\label{fig3}  a) The effect of decoherence sources $\eta_1=1/2T_{f,1}$ on the fidelity of state  transfer operation $\ket{\uparrow0}\rightarrow-i\ket{\downarrow1}$ with different non-JC couplings $g'$ (from top to bottom: $g'/g=0$,1,2,3,4,5,6.). Other parameters are as in Fig.\ref{fig2}. b) The same plot for the the influence of $\eta_1=1/T_{f,2}$. }
\end{figure}

 To sufficiently suppress the influence of the non-JC coupling, $g/g'\geq1/3$ is required (as we explain later in detail), which may be fulfilled by choosing $\beta\gg1$, $\alpha\rightarrow0.5$, e.g., we have
$g/g'\approx2$ for the case where $\beta=15$, $\alpha=0.8$, $E_J/E_C=80$.  The corresponding flux quantum fluctuation is $\zeta=0.14$ and the  phase difference of Josephson junction $j_4$ is $\theta=0.05$, which is within the reach of  a phase controller \cite{ljclk}. Apart from the non-JC coupling, there are other relevant imperfections for the hybrid system. The tunneling between $\ket{0}_f$ and $\ket{1}_f$ with tunneling rate $r\sim\omega_f \text{exp}(-\sqrt{E_J/E_C})$ decreases the coherence time of the superconducting flux qubit. The coupling strength $g$ should be strong enough to repress the unwanted tunneling probability $(r/g)^2$ \cite{ljck}. Low temperature is required to exponentially decrease the probability of the occupation of the excitation modes of the quantum wire by the factor $\text{exp}(\frac{-v_f}{k_BTL})$ \cite{ljck}.

 \section{Numerical simulations}
 Considering the decoherence sources, the dynamical process of the hybrid system is described by the Lindblad master equation
\bea\label{eq7}
\frac{\partial\rho}{\partial t}&=&-i[H_I,\rho]+\frac{1}{2T_{f,1}}(2a\rho a^\dagger-a^\dagger a\rho-\rho a^\dagger a)\notag\\
&+&\frac{1}{T_{f,2}}(\sigma_f^z\rho\sigma_f^z-\rho)
\eea
where the decoherence from the topological qubit has been neglected due to this qubit's great merit of long coherence time, $T_{f,1}$ and $T_{f,2}$ are the relaxation time and dephasing time of the superconducting flux qubit, respectively \cite{ymgs}.

To study the quantum information transfer between the topological and the superconducting flux qubits under realistic conditions we numerically simulate the master equation \eqref{eq7}. We may set $\alpha=0.8$, $\beta=15$, $E_J/E_C=80$, $E_J/2\pi=158$GHz, $\omega_f/2\pi=50$GHz, $T_{f,1}=900$ns, and $T_{f,2}=20$ns  for the superconducting flux qubit \cite{ljck,icyn}; the parameters for the topological qubit may assume to be $\Delta_0/2\pi=32.5$ GHz, $v_F=10^5$m/s, $L=5\mu$m \cite{ljck,vmou}.  The resonance condition gives $E(\phi_{\text{on}})/2\pi=\omega_f/2\pi=50$GHz, resulting in $\phi_{\text{on}}=-1.73$ according to \eqref{eq2} with $\Lambda_{\phi_{\text{on}}}=-7.75 $. Then equations (\ref{eq10}) give the the coupling strength   $g/2\pi=-2$ GHz and  $g'/2\pi=-1$ GHz.
The evolution of the state transfer
\be\label{eq8}
\ket{\uparrow0}\xrightarrow{\int^{t_{f1}} g(t)dt=-\pi}\ket{\psi_1}\equiv-i\ket{\downarrow1}
 \ee
 and the generating of a maximally entangled state
 \be\label{eq9}
 \ket{\uparrow0}\xrightarrow{\int^{t_{f2}} g(t)dt=-\pi/2}\ket{\psi_2}\equiv(\ket{\uparrow0}-i\ket{\downarrow1})/\sqrt{2}
  \ee
  are shown in Fig. \ref{fig2}a) and b), respectively, with the corresponding fidelity $F_1=\bra{\psi_1}\rho(t_{f1})\ket{\psi_1}=0.993$ and $F_2=\bra{\psi_2}\rho(t_{f2})\ket{\psi_2}=0.996$. Fig.\ref{fig3} shows the influence of the decoherence sources $\eta_1=1/2T_{f,1}$, $\eta_2=1/T_{f,2}$, and $g'$ on the state transfer fidelity $F_1$. From Fig.\ref{fig3} we see that the influence of $g'$ on the state transfer is small: $F_1=0.982$ for the case where $g'=3g=-6(2\pi)$GHz, $E_(\phi_{\text{on}})$, $\omega_f$, $T_{f,1}$, $T_{f,2}$, $v_F$, $L$, and $\Lambda_{\phi_{\text{on}}} $  remain  the same as in fig.\ref{fig2}, while other parameters are $\alpha=0.97$, $\beta=10$, $E_J/E_C=30000$ \cite{jmsn, ljclk}, $\theta=0.086$, $\zeta=0.04$, $E_J=3.1$ THz, $\phi_{\text{on}}=-0.646$, and $\Delta_0=78$GHz.

Apart from the aforesaid decoherence sources, there exist processes which may influence the interaction between the two Majorana fermions, such as dynamic modulations of the superconducting gap and variation of the electromagnetic environment owing to charge fluctuations.  We estimate their influence on the operation fidelity by assuming unknown errors in $E(\phi_{\text{on}})$, $g'$, and $g$, and find that the corresponding fidelity $F_1$ decreases from 0.993 to 0.968 for even 10\% unknown errors in $E(\phi_{\text{on}})$, $g'$, and $g$.

The recent proposal \cite{ljck} applies in the parameter regime $g'\gg g$, in contrast our scheme works well in the parameter regime $g'\leq 3 g$. With Jaynes-Cummings coupling quantum state transfer and quantum entanglement distribution between the topological and flux qubits can be more conveniently accomplished.

\section{CONCLUSIONS}
 In summary, we have presented a scheme for quantum information transfer between topological and superconducting flux qubits. A strong Jaynes-Cummings coupling between topological and flux qubits is achieved. With this scheme, quantum state transfer, quantum entanglement generating, and arbitrary unitary transformation in the topological-flux hybrid system may be accomplished with near unit fidelity. This quantum interface enable us to store quantum information on topological qubits for long-time storage, to efficiently read out of topological qubit states, to implement partially protected universal topological quantum computation, where single-qubit state of flux qubit can be  prepared with high accuracy and is transferred to topological qubit to compensate topological qubit's incapability of generating some single-qubit states.

%\begin{acknowledgments}
 This work was supported by the National Natural Science Foundation of China ( 11072218 and 11272287), by Zhejiang Provincial Natural Science Foundation of China (Grant No. Y6110314), and  by Scientific Research Fund of Zhejiang Provincial Education Department (Grant No. Y200909693).
%\end{acknowledgments}
%%%%%%%%%%%% Bibliography %%%%%%%%%%%%%%%%%%%%%%%%


\begin{references}
\bibitem{ayki}A.Y. Kitaev,  Ann. Phys. (N.Y.) {\bf 303}, 2 (2003).
\bibitem{cnsh}C. Nayak, S.H. Simon, A. Stern, M. Freedman, and S. Das
Sarma, Rev. Mod. Phys. {\bf 80}, 1083 (2008).
\bibitem{fwil}F. Wilczek, Nature Phys {\bf 5}, 614 (2009).
\bibitem{nrdg}N. Read and D. Green,  Phys. Rev. B {\bf 61}, 10267 (2000).
\bibitem{sscn}S. Das Sarma, C. Nayak, and S. Tewari,  Phys. Rev.B {\bf 73}, 220502R (2006).
\bibitem{lfck}L. Fu and C.L. Kane,  Phys. Rev. Lett. {\bf 100}, 096407 (2008).
\bibitem{mhck}M. Z. Hasan and C. L. Kane,  Rev. Mod. Phys. {\bf 82}, 3045(2010).
\bibitem{jsrl}J. D. Sau, R. M. Lutchyn, S. Tewari, and S. Das Sarma,  Phys. Rev. Lett. {\bf 104}, 040502(2010).
\bibitem{jali}J. Alicea,  Phys. Rev. B {\bf 81}, 125318 (2010).
\bibitem{yogr}Y. Oreg, G. Refael, and F. von Oppen, Phys. Rev. Lett. {\bf 105}, 177002 (2010).
\bibitem{rljs}R. M. Lutchyn, J. D. Sau, and S. Das Sarma,  Phys. Rev. Lett. {\bf 105}, 077001 (2010).
\bibitem{vmou}V. Mourik, K. Zuo, S. M. Frolov, S. R. Plissard, E. P. A. M. Bakkers, L. P. Kouwenhoven, Science {\bf 336}, 1003 (2012).
\bibitem{daiv}D. A. Ivanov,  Phys. Rev. Lett. {\bf 86}, 268 (2001).
\bibitem{aste}A. Stern,  Nature( London) {\bf 464}, 187 (2010).
\bibitem{pbon}P. Bonderson,  Phys. Rev. Lett. {\bf 103}, 110403 (2009).
\bibitem{pbond}P. Bonderson, D.J. Clarke, C. Nayak, and K. Shtengel,  Phys. Rev. Lett. {\bf 104}, 180505(2010).
\bibitem{pbrl}P. Bonderson and R.M. Lutchyn, Phys. Rev. Lett. {\bf 106}, 130505 (2011).
\bibitem{srcn}S. Ritter, C. N\"{o}lleke, C. Hahn, A. Reiserer, A. Neuzner, M. Uphoff, M. M\"{u}cke, E. Figueroa, J. Bochmann, and  G. Rempe, Nature (London) {\bf 484}, 195 (2012).
\bibitem{mglc}M.V. Gurudev Dutt, L. Childress, L. Jiang, E. Togan, J. Maze, F. Jelezko, A.S. Zibrov, P.R. Hemmer, and M.D. Lukin, Science {\bf 316}, 1312 (2007).
\bibitem{pmgk}P.C. Maurer, G. Kucsko, C. Latta, L. Jiang, N.Y. Yao, S.D. Bennett, F. Pastawski,
D. Hunger, N. Chisholm, M. Markham, D.J. Twitchen, J.I. Cirac, and M.D. Lukin, Science {\bf 336}, 1283 (2012).
\bibitem{rbdw}R. Blatt and D. Wineland,  Nature (London) {\bf 453}, 1008 (2008).
\bibitem{jcfw}J. Clarke and F.K. Wilhelm,  Nature (London) {\bf 453}, 1031 (2008).










\bibitem{dmpm}D.L. Moehring, P. Maunz, S. Olmschenk, K.C. Younge, D. N. Matsukevich, L.-M. Duan, and  C. Monroe,  Nature (London) {\bf 449}, 68 (2007).
\bibitem{etyc}E. Togan, Y. Chu, A.S. Trifonov, L. Jiang, J. Maze, L. Childress, M.V.G. Dutt, A.S. S{\o}ensen, P.R. Hemmer, A.S. Zibrov, and  M.D. Lukin,  Nature (London) {\bf 466}, 730 (2010).








\bibitem{ljia}L. Jiang, G.K. Brennen, A.V. Gorshkov, K. Hammerer, M. Hafezi, E. Demler, M.D. Lukin, and P. Zoller,  Nature Phys. {\bf 4}, 482 (2008).
\bibitem{magu}M. Aguado, G. K. Brennen, F. Verstraete, and J. I. Cirac,  Phys. Rev. Lett. {\bf 101}, 260501 (2008).
\bibitem{fhaa}F. Hassler, A.R. Akhmerov, C.-Y. Hou, and C.W. J.
      Beenakker, New J. Phys. {\bf 12}, 125002 (2010).
\bibitem{jsst}J.D. Sau, S. Tewari, and S. Das Sarma,  Phys. Rev. A {\bf 82}, 052322 (2010).
\bibitem{ljck}L. Jiang, C.L. Kane, and J. Preskill,  Phys. Rev. Lett. {\bf 106}, 130504 (2011).
\bibitem{jemo}J.E. Mooij, T.P. Orlando, L. Levitov, L. Tian, C.H. van der Wal, and S. Lloyd, Science {\bf 285}, 1036 (1999).
  \bibitem{zxss}X. Zhu, S. Saito, A. Kemp, K. Kakuyanagi, S. Karimoto, H. Nakano, W.J. Munro,
Y. Tokura, M.S. Everitt, K. Nemoto, M. Kasu, N. Mizuochi, and  K. Semba,  Nature( London) {\bf 478}, 221 (2011).
 \bibitem{ljclk}L. Jiang, C. L. Kane, and J. Preskill,  arXiv: 1010.5862v2.



\bibitem{mnic}M.A. Nielsen and I.L. Chuang,  Quantum Computation and Quantum Information, (Cambridge University Press, Cambridge, England, 2010).




\bibitem{ymgs}Y. Makhlin, G. Sch\"{}n, and A. Shnirman, Rev. Mod. Phys. {\bf 73}, 357 (2001).

\bibitem{icyn}I. Chiorescu, Y. Nakamura, C.J.P.M. Harmans, and J.E. Mooij,  Science {\bf 299}, 1869 (2003).

\bibitem{jmsn}J. M. Martinis, S. Nam, J. Aumentado, and C. Urbina,  Phys. Rev. Lett. {\bf 89}, 117901(2002).




















\end{references}
\end{document}